\documentclass{cernphprep}

\usepackage{times}
\usepackage{graphicx}
\usepackage{bm}

\usepackage{textcomp} 
\usepackage{color}

\newcommand{\app}{\mbox{$A_{2\pi}$}}

\def\pipi{\mbox{$\pi^+\pi^-$}}

\begin{document}

\begin{titlepage}

\EXPnumber{DIRAC/PS212}
\PHnumber{2015--175}
\PHdate{\today}

\title{First observation of long-lived $\pi^+ \pi^-$ atoms}

\begin{Authlist}

B.~Adeva\Iref{s}, 
L.~Afanasyev\Iref{d}, 
A.~Anania\Iref{im},
S.~Aogaki\Iref{b},
A.~Benelli\Iref{cz}, 
V.~Brekhovskikh\Iref{p},  
T.~Cechak\Iref{cz}, 
M.~Chiba\Iref{jt}, 
P.~Chliapnikov\Iref{p},  
P.~Doskarova\Iref{cz}, 
D.~Drijard\Iref{c},
A.~Dudarev\Iref{d},
M.~Duma\Iref{b},  
D.~Dumitriu\Iref{b}, 
D.~Fluerasu\Iref{b}, 
A.~Gorin\Iref{p}, 
O.~Gorchakov\Iref{d},
K.~Gritsay\Iref{d}, 
C.~Guaraldo\Iref{if}, 
M.~Gugiu\Iref{b}, 
M.~Hansroul\Iref{c}, 
Z.~Hons\Iref{czr}, 
S.~Horikawa\Iref{zu},
Y.~Iwashita\Iref{jk},
V.~Karpukhin\Iref{d}, 
J.~Kluson\Iref{cz}, 
M.~Kobayashi\Iref{k}, 
V.~Kruglov\Iref{d}, 
L.~Kruglova\Iref{d}, 
A.~Kulikov\Iref{d}, 
E.~Kulish\Iref{d},
A.~Kuptsov\Iref{d}, 
A.~Lamberto\Iref{im}, 
A.~Lanaro\Iref{u},
R.~Lednicky\Iref{cza}, 
C.~Mari\~nas\Iref{s},
J.~Martincik\Iref{cz},
L.~Nemenov\IIref{d}{c},
M.~Nikitin\Iref{d}, 
K.~Okada\Iref{jks}, 
V.~Olchevskii\Iref{d},
V.~Ovsiannikov\Iref{v},
M.~Pentia\Iref{b}, 
A.~Penzo\Iref{it}, 
M.~Plo\Iref{s}, 
P.~Prusa\Iref{cz},  
G.~Rappazzo\Iref{im}, 
A.~Romero Vidal\Iref{s},
A.~Ryazantsev\Iref{p},
V.~Rykalin\Iref{p},
J.~Saborido\Iref{s}, 
J.~Schacher\IAref{be}{*},
A.~Sidorov\Iref{p}, 
J.~Smolik\Iref{cz}, 
F.~Takeutchi\Iref{jks}, 
L.~Tauscher\Iref{ba},
T.~Trojek\Iref{cz}, 
S.~Trusov\Iref{m}, 
T.~Urban\Iref{cz},
T.~Vrba\Iref{cz},
V.~Yazkov\Iref{m}, 
Y.~Yoshimura\Iref{k}, 
M.~Zhabitsky\Iref{d}, 
P.~Zrelov\Iref{d} 

\end{Authlist}

\Instfoot{s}{Santiago de Compostela University, Spain}
\Instfoot{d}{JINR Dubna, Russia}
\Instfoot{im}{INFN, Sezione di Trieste and Messina University, Messina, Italy}
\Instfoot{b}{
IFIN-HH, National Institute for Physics and Nuclear Engineering, 
Bucharest, Romania
}
\Instfoot{cz}{Czech Technical University in Prague, Czech Republic}
\Instfoot{p}{IHEP Protvino, Russia}
\Instfoot{jt}{Tokyo Metropolitan University, Japan}
\Instfoot{c}{CERN, Geneva, Switzerland}
\Instfoot{if}{INFN, Laboratori Nazionali di Frascati, Frascati, Italy}
\Instfoot{czr}{Nuclear Physics Institute ASCR, Rez, Czech Republic}
\Instfoot{zu}{Zurich University, Switzerland}
\Instfoot{jk}{Kyoto University, Kyoto, Japan}
\Instfoot{k}{KEK, Tsukuba, Japan}
\Instfoot{u}{University of Wisconsin, Madison, USA} 
\Instfoot{cza}{Institute of Physics ASCR, Prague, Czech Republic}
\Instfoot{jks}{Kyoto Sangyo University, Kyoto, Japan}
\Instfoot{v}{Voronezh State University, Russia}
\Instfoot{it}{INFN, Sezione di Trieste, Trieste, Italy}
\Instfoot{be}{
Albert Einstein Center for Fundamental Physics, 
Laboratory of High Energy Physics, Bern, Switzerland
}
\Instfoot{ba}{Basel University, Switzerland}
\Instfoot{m}{
Skobeltsin Institute for Nuclear Physics of Moscow State University, 
Moscow, Russia
}

\Anotfoot{*}{Corresponding author} 

\Collaboration{DIRAC Collaboration}
\ShortAuthor{DIRAC Collaboration}

\newpage

\begin{abstract}
After observing and investigating the double-exotic $\pi^+\pi^-$ atom 
with the ground state lifetime $\tau$ of about $3 \times 10^{-15}$~s, 
the upgraded DIRAC experiment at the CERN PS accelerator observes 
for the first time long-lived states of the same atom with lifetimes 
of about $10^{-11}$~s and more. The number of characteristic 
pion pairs resulting from the breakup (ionisation) of long-lived 
$\pi^+\pi^-$ atoms amounts to $436\pm61$, corresponding to 
a signal-to-error ratio of better than 7 standard deviations. 
This observation opens a new possibility to measure energy differences 
between $p$ and $s$ atomic states and so to determine 
$\pi \pi$ scattering lengths. 
\end{abstract}
\vspace{2cm}
\Submitted{(To be submitted to Physics Letters B)}
\end{titlepage}

\section{Introduction}
\label{sec:intro}
 
The experimental program of the DIRAC collaboration comprises 
the observation and detailed analysis of dimesonic atoms, 
which are produced by protons interacting 
with target nuclei \cite{NEME85}. 
First, DIRAC has investigated $\pi^+\pi^-$ atoms (pionium, \app{}) 
and measured their lifetime, $\tau=(3.15_{-0.26}^{+0.28})$~fs, 
in the ground state and hence a combination 
of $\pi \pi$ scattering lengths \cite{ADEV11}. 
Second, evidence for the production of dimesonic atoms with 
strangeness, i.e. $\pi K$ atoms, has been found \cite{ADEV09}.
The data analysis has yielded a first measurement of 
the $\pi K$ atom lifetime and  
$\pi K$ scattering lengths \cite{ADEV14}. 
Third, a search for long-lived $\pi \pi$ atom states has been 
performed. DIRAC plans to study the \textit{Lamb shift} in 
$\pi \pi$ atoms and then to extract 
another $\pi \pi$ scattering length combination. 
In this paper DIRAC presents the first observation of more than  
400 long-lived $\pi^+\pi^-$ atoms.  

The decay probability of \pipi{} atoms is dominated by the
annihilation process
\begin{equation}\label{eq.anh}
\pi^{+} + \pi^{-} \rightarrow \pi^{0} + \pi^{0}
\end{equation}
(branching ratio $\sim 99.6\%$) and depends on the
difference between the $S$-wave $\pi\pi$ scattering lengths 
with isospins zero ($a_0$) and two ($a_2$)
\cite{Uretsky61,Bilenky69,Jallouli98,Ivanov98}:
\begin{equation}\label{eq:tau-aa}
\frac{1}{\tau} \approx W_{\pi^0\pi^0} = 
R \left| a_0-a_2 \right|^2 \quad \mbox{with} \quad 
R \propto \left|\psi_{nl}(0)\right|^2.
\end{equation}

The expression $\psi_{nl}(0)$ is the pure Coulomb atomic 
wave function at the origin with principal quantum number $n$ 
and angular momentum quantum number $l$. The most accurate ratio 
$R$ has been derived with a precision of 1.2\% in \cite{Gasser01}. 

In order to get values of $a_0$ and $a_2$ separately from 
$\pi^+\pi^-$ atom data, one may exploit the fact 
that the energy splitting between the levels $ns$ and $np$, 
$\Delta E^{n\mathrm{s}-n\mathrm{p}}=~E_{ns}-E_{np}$, 
depends on another combination of the scattering lengths: 
$2a_0+a_2$ \cite{EFIM86}. The influence of 
strong and electromagnetic interactions 
on the \app{} energy structure has been studied in 
\cite{EFIM86,KARI79,AUST83,GASH98,EIRA00} and 
a detailed analysis performed in \cite{SCHW04}.
The energy shift for the levels with the principal quantum number $n$ 
and orbital quantum number $l$ is composed of three contributions:
$\Delta E_{nl}=~\Delta E_{nl}^{\mathrm{em}} + \Delta E_{nl}^{\mathrm{vac}} 
 + \Delta E_{nl}^{\mathrm{str}}$. The term 
$\Delta E_{nl}^{\mathrm{em}}$ includes relativistic insertions,
finite-size effect, self-energy corrections due to transverse photons
and transverse photon exchange. Contributions from vacuum polarization 
are covered by $\Delta E_{nl}^{\mathrm{vac}}$. The last term 
$\Delta E_{nl}^{\mathrm{str}}$ takes into account effects from 
strong interaction and is related to the $\pi\pi$ scattering lengths 
according to: 
$\Delta E_{n0}^{\mathrm{str}}=A_{n}(2a_0+a_2)$.
The theoretical value for the $2\mathrm{s}-2\mathrm{p}$ energy splitting amounts 
to $\Delta E^{2\mathrm{s}-2\mathrm{p}}=(-0.59 \pm 0.01)~\mathrm{eV}$, 
whereas the splitting $\Delta E^{n\mathrm{s}-n\mathrm{p}}$ for 
higher principal quantum numbers $n$ decreases \cite{SCHW04}. 
By measuring the value of $\Delta E^{n\mathrm{s}-n\mathrm{p}}$,  
one can determine the numerical value of
$\Delta E_{n0}^{\mathrm{str}}$, as all other terms in $\Delta E_{nl}$ 
have been calculated with high precision: 
the strong interaction effects contribute up 
to 80\% of the full energy shift. This fact provides 
a high sensitivity of a $\Delta E^{2\mathrm{s}-2\mathrm{p}}$ measurement
to the value of the term $2a_0+a_2$. Thus, a measurement of 
the energy shift $\Delta E^{n\mathrm{s}-n\mathrm{p}}$  
allows to obtain a value for the new combination of 
scattering lengths $2a_0+a_2$. 

A method to measure $\Delta E^{n\mathrm{s}-n\mathrm{p}}$ 
has been proposed in \cite{NEME85}. By investigating 
the influence of an applied electric or magnetic field on 
the decay probability of long-lived 
$\pi^+\pi^-$ atom states $A_{2\pi}^L$ ($l\geq1$),  
it is possible to extract an experimental value for 
$\Delta E^{n\mathrm{s}-n\mathrm{p}}$~\cite{NEME01,NEME02}.   

\begin{figure}[ht]
  \centering
  \includegraphics[width=100mm]{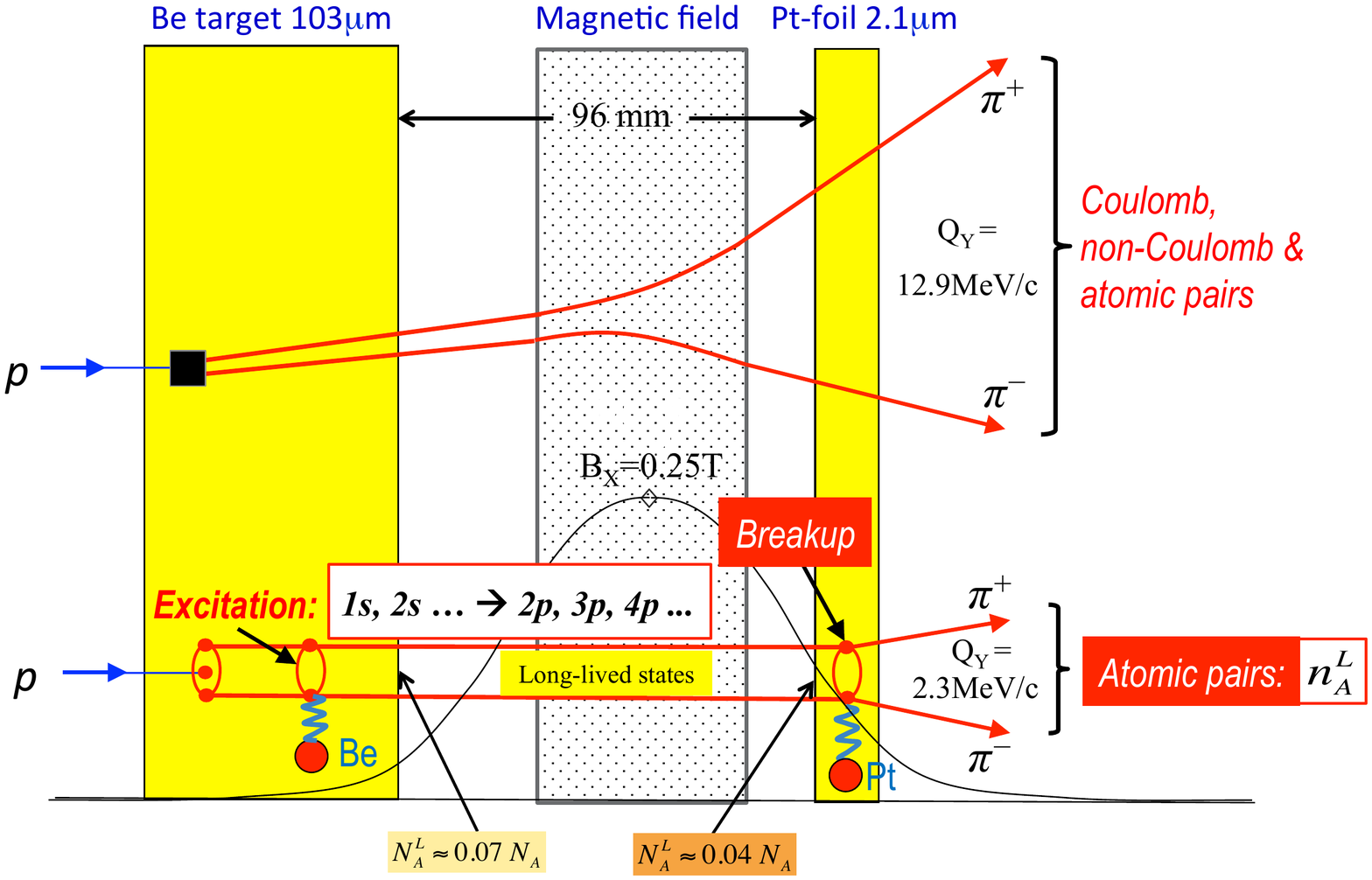}
  \caption{Method to observe long-lived $A_{2\pi}^L$ by means of 
               a breakup foil (Pt).}
  \label{Fig1_1}
\end{figure}

In inclusive processes \app{} atoms are produced in $s$-states 
distributed over the principal quantum number $n$ according to 
$n^{-3}$. When moving inside the target, relativistic \app{}  
interact with the electric field of target atoms, 
and some of them ($N_A^L$) leave the target 
with an orbital quantum number $l>0$.  
The main primary excitation process is the transition 
$ns \rightarrow n'p$. Then, for excited states with $l \geq 1$, 
the decays of $A_{2\pi}^L$ into two $\pi^0$, 
$\pi^0 + \gamma$ and two $\gamma$ are suppressed in 
accordance with (\ref{eq:tau-aa}) because of 
$\left|\psi_{nl}(0)\right|^2=0$~\cite{NEME01}. 
Therefore, the decay mechanism of such excited states 
is the radiative deexcitation to an $ns$ state, annihilating 
subsequently with the lifetime $\tau \cdot n^3$ 
into two $\pi^0$. Thus, the $A_{2\pi}^L$ decay probability 
is given by the shortest radiative lifetime, 
the $2p$ lifetime $\tau_{2p}=1.17 \cdot 10^{-11}$~s. 
For an average \app{} momentum of 4.5~GeV/$c$ 
($\gamma \simeq 16$), the decay lengths are 
5.7~cm ($2p$), 19~cm ($3p$) and 43~cm ($4p$).
Using a $\sim$100~$\mu$m thick Be target and inserting 
a $\sim$2~$\mu$m thick Pt foil downstream of 
this target \cite{SPSC11}, a large fraction of 
the long-lived atoms $A_{2\pi}^L$, generated in Be, 
reaches the Pt foil and breaks up, thus  
providing an extra number $n_A^L$ of atomic pairs 
(see Fig.~\ref{Fig1_1}).

In order to be able to claim ``observation of 
long-lived $A_{2\pi}^L$'' -- the goal of 
this investigation -- the DIRAC setup has been modified 
in 2012 as described in the following section.  

The most accurate theoretical predictions for the $S$-wave 
$\pi\pi$ scattering lengths have been achieved in \cite{COLA01} 
(in units $M_{\pi^+}^{-1}$):
\begin{equation}
  \label{eq:a0a2}
  a_0=0.220\pm0.005\,, \quad a_2=-0.0444\pm0.0010\,, \quad
  a_0-a_2=0.265\pm0.004 \;.
\end{equation}
The best experimental results with a precision of around 
4\% have been obtained from studying the decays 
$K^\pm \to \pi^+ \pi^- e^\pm \nu$ \cite{BATL10} and 
$K^\pm \to \pi^\pm \pi^0 \pi^0$ \cite{BATL09} 
as well as from measuring 
the $\pi^+\pi^-$ atom lifetime \cite{ADEV11}: 
 $\left|a_{0}-a_{2}\right|=
\left( 0.2533_{-0.0107}^{+0.0112} \right)$. 
In the case of $K$ decays, additional theoretical information 
can improve the experimental scattering length values 
\cite{BATL10,BATL09}.

\section{Setup for detection of long-lived $\pi^+\pi^-$ atoms}
\label{sec:setup}
	 
\begin{figure}[ht]
\begin{center}
\includegraphics[width=100mm]{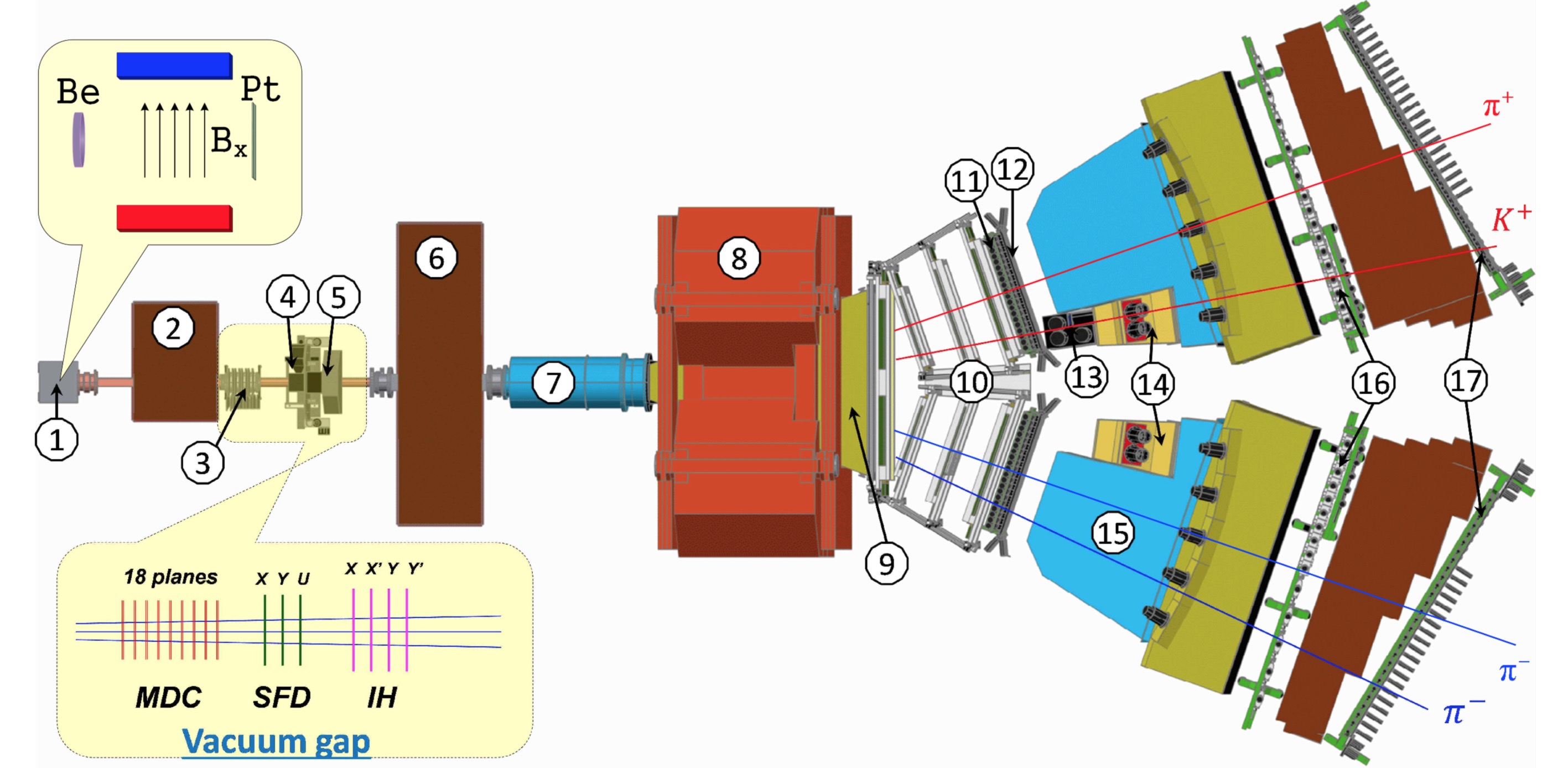}
\caption{General view of the DIRAC setup:  
1 -- target station with insertion, showing the Be target, 
magnetic field and Pt breakup foil;
2 -- first shielding;
3 -- microdrift chambers (MDC);
4 -- scintillating fiber detector (SFD); 
5 -- ionisation hodoscope (IH); 
6 -- second shielding; 
7 -- vacuum tube; 
8 -- spectrometer magnet; 
9 -- vacuum chamber; 
10 -- drift chambers (DC); 
11 -- vertical hodoscope (VH); 
12 -- horizontal hodoscope (HH); 
13 -- aerogel Cherenkov; 
14 -- heavy gas Cherenkov; 
15 -- nitrogen Cherenkov; 
16 -- preshower (PSh); 
17 -- muon detector. 
(The plotted symmetric and asymmetric events are 
a $\pi \pi$ and $\pi K$ pair, respectively.)}
\label{Fig2_1}
\end{center}
\end{figure}

The DIRAC setup \cite{DIRA15} is sketched in Fig.~\ref{Fig2_1}.
A high-resolution magnetic spectrometer 
$(\Delta p/p \simeq 3\cdot 10^{-3})$  
is used to split oppositely charged meson pairs 
($\pi^+ \pi^-$ and $\pi^\mp K^\pm$) and 
to measure their relative c.m. momentum ($Q$) 
with good precision in order to extract 
a dimeson atom signal \cite{ADEV11,ADEV14}. 
The 24~GeV/c CERN PS proton beam interacting with the target 
produces additionally free (unbound) ``Coulomb pairs'' 
from short-lived resonances, ``non-Coulomb pairs'' 
from long-lived sources and accidental coincidences 
(different proton-nucleus interactions). 
Therefore, the atom signal suffers from a $\pi^+\pi^-$ 
continuum background causing the main signal uncertainty.

\begin{figure}[ht]
  \centering
  \includegraphics[width=100mm]{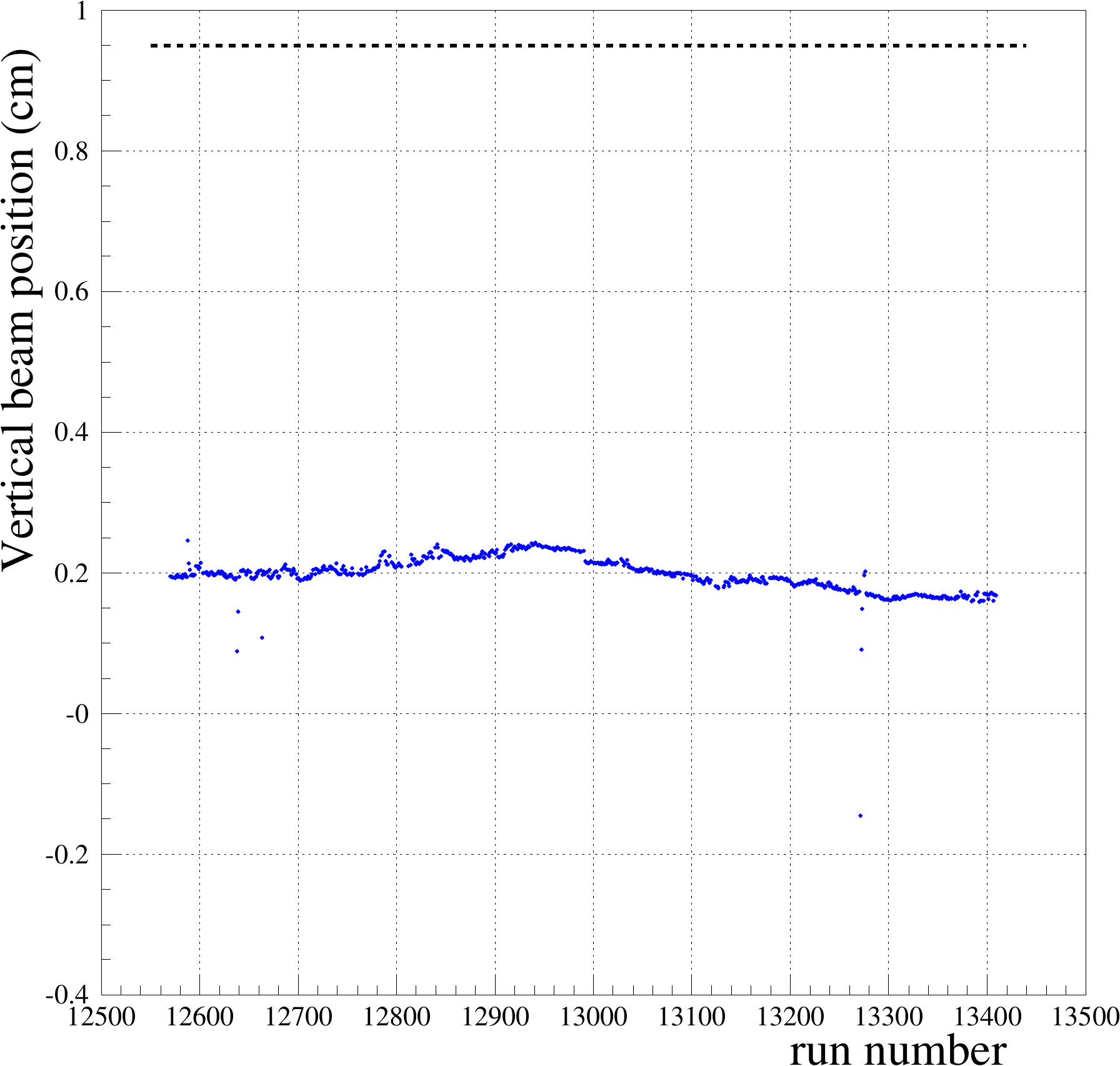}
  \caption{Vertical coordinate of the beam position (dots)  
           on the Beryllium target, reconstructed with 
	   track information for all data runs. 
           The low edge of the Platinum foil is shown 
	   as dashed line.}
  \label{Fig2_2}
\end{figure}

In the search for long-lived $A_{2\pi}^L$,  
the primary proton beam hits 
a 103 $\mu$m thick 99.98\% pure Be target, 
providing the needed $A_{2\pi}^L$ yield at 
an acceptable proton beam intensity~\cite{SPSC11}. 
The target radiation thickness amounts to 
$3.0 \cdot 10^{-4}$~$X_{0}$ (radiation length) and 
the nuclear interaction probability to 
$\epsilon_{nuc} = 2.5 \cdot 10^{-4}$. 
The secondary channel with the whole setup is  
vertically inclined relative to the proton beam 
by $5.7^\circ$ upward. By means of 
a rectangular beam collimator inside of  
the second steel shielding wall (Fig.~\ref{Fig2_1}, item 6), 
secondary particles are confined to $\pm 1^\circ$ 
in the horizontal (X) and vertical (Y) planes and thus 
to the solid angle $\Omega = 1.2 \cdot 10^{-3}$~sr. 
Downstream of the Be target, a 2.1 $\mu$m thick Pt foil 
($6.9 \cdot 10^{-4}$~$X_{0}$,  
$\epsilon_{nuc} = 0.23 \cdot 10^{-4}$) for 
$A_{2\pi}^L$ breakup has been placed at a distance of 96~mm. 
The foil is installed at 7.5~mm above the primary proton beam 
to avoid interaction of the beam halo with Pt (Fig.~\ref{Fig1_1}). 
The upper limit of the beam size in the vertical direction is 
$\sigma_y = 1.75$~mm \cite{note1502}. 
The beam position in the vertical plane 
has been permanently monitored during the run by checking 
SFD and IH counting rates and by reconstructing 
the beam position with track information (Fig.~\ref{Fig2_2}). 
Between target and breakup foil, a permanent 
magnet \cite{note1304} has been introduced 
to suppress significantly background 
$\pi^+\pi^-$ pairs, generated in Be, 
in the low $Q$ region, where the atomic pairs from 
$A_{2\pi}^L$ breakup are expected. 
The retractable magnet with a pole distance of 60~mm consists 
of a Samarium-Cobalt alloy ($\text{Sm}_2 \text{Co}_{17}$) and 
has a maximum horizontal field strength of 0.25~T 
(see insertion in Fig.~\ref{Fig2_1}). 
The bending power of 0.02~Tm is in the secondary beam region 
relatively homogenous with a precision of better than $\pm$ 2\%.

\begin{figure}[ht]
  \centering
  \includegraphics[width=75mm]{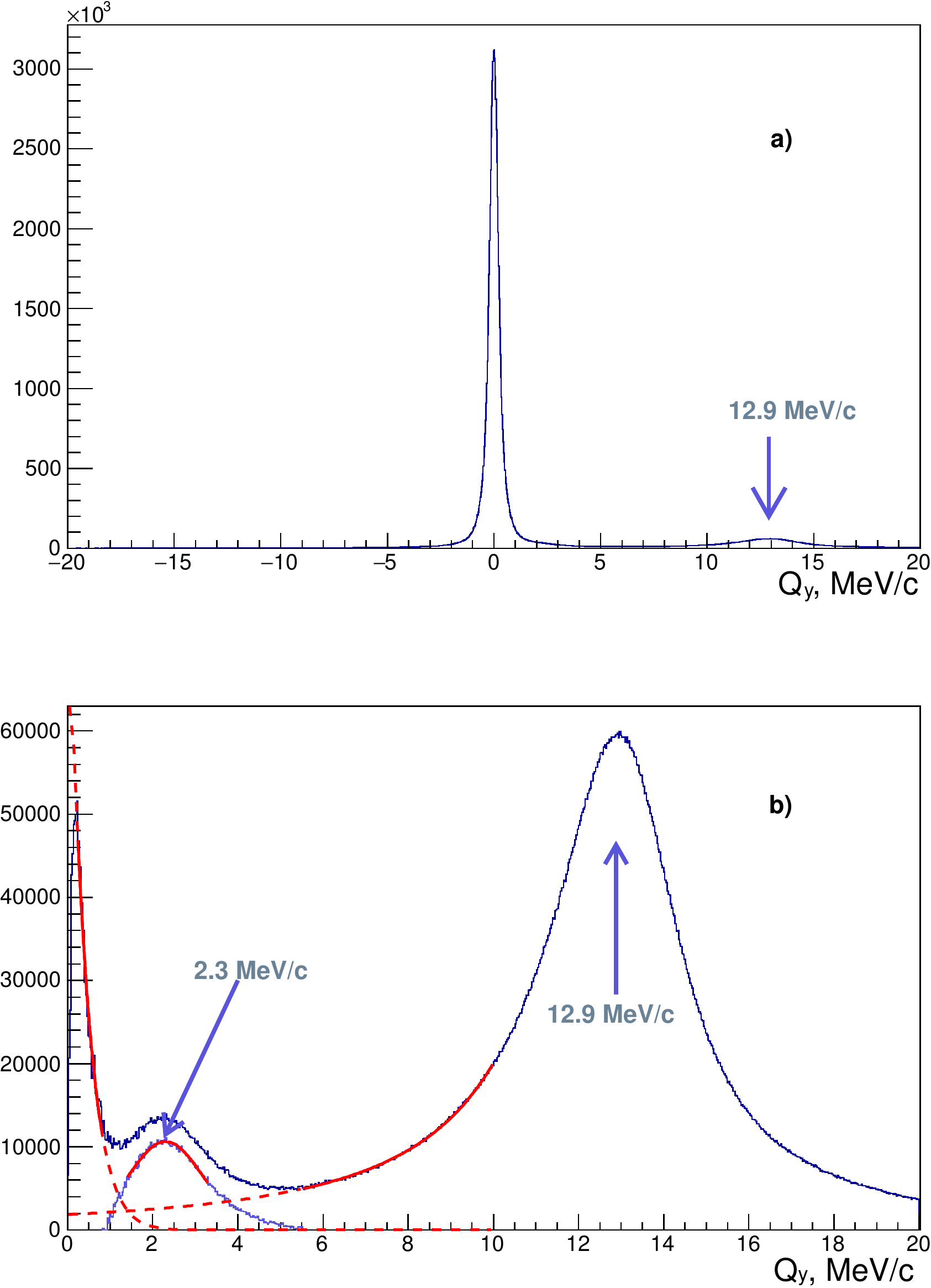}
  \caption{a) Experimental distribution 
   of e$^+$e$^-$ pairs on $Q_Y$. 
   b) Same distribution after mirror subtraction.}
  \label{Fig2_3}
\end{figure}

In order to measure the shift of the vertical component $Q_Y$, 
which is enlarged by the horizontal magnetic field, 
$\mathrm{e}^+\mathrm{e}^-$ Dalitz pairs generated in the Be target 
as well as  $\mathrm{e}^+\mathrm{e}^-$ pairs 
produced in the Pt foil by photons have been investigated. 
The experimental $Q_Y$ distribution (Fig.~\ref{Fig2_3}a) 
shows a first peak around  originating from pairs 
produced far downstream of the magnet and a second peak from 
Dalitz pairs at $Q_Y=12.9$~MeV/$c$. After subtracting 
the left part of the central peak ($Q_Y \approx 0$) from 
the right part (mirrored subtraction), a peak at 
$Q_Y=2.3$~MeV/$c$ appears from $\mathrm{e}^+\mathrm{e}^-$ pairs 
produced in Pt and crossing only the fringing magnetic 
field \cite{note1402} (Fig.~\ref{Fig2_3}b). 
The experimental peak positions as well as the shapes 
coincide with simulation, which has been obtained by  
using the magnetic field map with a precision of better 
than 1\% \cite{note1403}. 
Each oppositely charged particle pair, 
generated in the Be target and hence crossing 
the magnetic field region, obtains an additional  
$\Delta Q_Y$ of 12.9~MeV/$c$, whereas a pair produced 
in the Pt foil gets a smaller $\Delta Q_Y$, 
caused by the fringing field, of only 2.3~MeV/$c$. 
This difference in the $Q_Y$ shift for pairs  
from Be and Pt allows to suppress significantly 
the background level for observing atomic pairs 
from long-lived $A_{2\pi}^L$. 

The peak position for Dalitz pairs at $Q_Y=12.9$~MeV/$c$ 
has been used to control the magnetic field stability 
during the 6 month data taking in 2012: 
the field strength of the $\text{Sm}_2 \text{Co}_{17}$  
magnet\footnote{Under the same radiation condition 
the field strength of a permanent Nd--Fe--B magnet 
has been decreased by more than 50\% during 
the 2011 run \cite{note1203}.}
has been stable within a relative precision of better 
than $5 \cdot 10^{-4}$ \cite{note1403}.

With a spill duration of 450~ms the beam intensity 
has been (28--30)$\cdot 10^{10}$ protons/spill.

\section{Event reconstruction}
\label{sec:track}

The event reconstruction has been performed by means of 
the DIRAC $\pi\pi$ analysis software already used for 
the analysis of the 2001--2003 data \cite{ADEV11}. 
Only events with one or two so-called DC tracks -- tracks 
reconstructed only with DC hits -- in each arm 
are processed according to the following criterions: \\
1) One or two hadron tracks are identified in the DC of 
both arms with corresponding hits in VH, HH and PSh, 
but no signal in ChN and Mu detectors (Fig.~\ref{Fig2_1}). 
In each arm the earliest track, 
which initiates the readout procedure, is selected for 
further analysis. \\
2) The DC tracks are extrapolated backward to 
the incident proton beam position on the target 
using the inverse transfer function of the DIRAC dipole magnet. 
This procedure provides a first approximation of 
the particle momenta and the corresponding 
intersection points in MDC, SFD and IH. \\
3) SFD hits are searched for in the coordinate region 
defined by the position accuracy: 
a square of size $\pm$1~cm, that corresponds to 
5$\sigma$ for high momentum and to 3$\sigma$ for 
low momentum particles. The two tracks should not use 
a common SFD hit in case of more than one hit in 
the proper region. If the number of hits around 
the two tracks is $\le$4 in each SFD plane and 
$\le$9 in all 3 SFD planes, the event is kept. These  
cuts reduce the data sample by 1/3, which is then  
called ``low and medium background events''. 
A further adjustment is applied in order to find 
the best two-track combination: 
The momenta of the positively and negatively 
charged particle are modified to match 
the DC track $X$-coordinates and the SFD hits 
in the $X$- or $U$-plane. In the final analysis, 
the combination with the best $\chi^2$ in 
the SFD planes and vertex in the target is selected. 

\begin{figure}[h] \centering{
\includegraphics[scale=0.5]{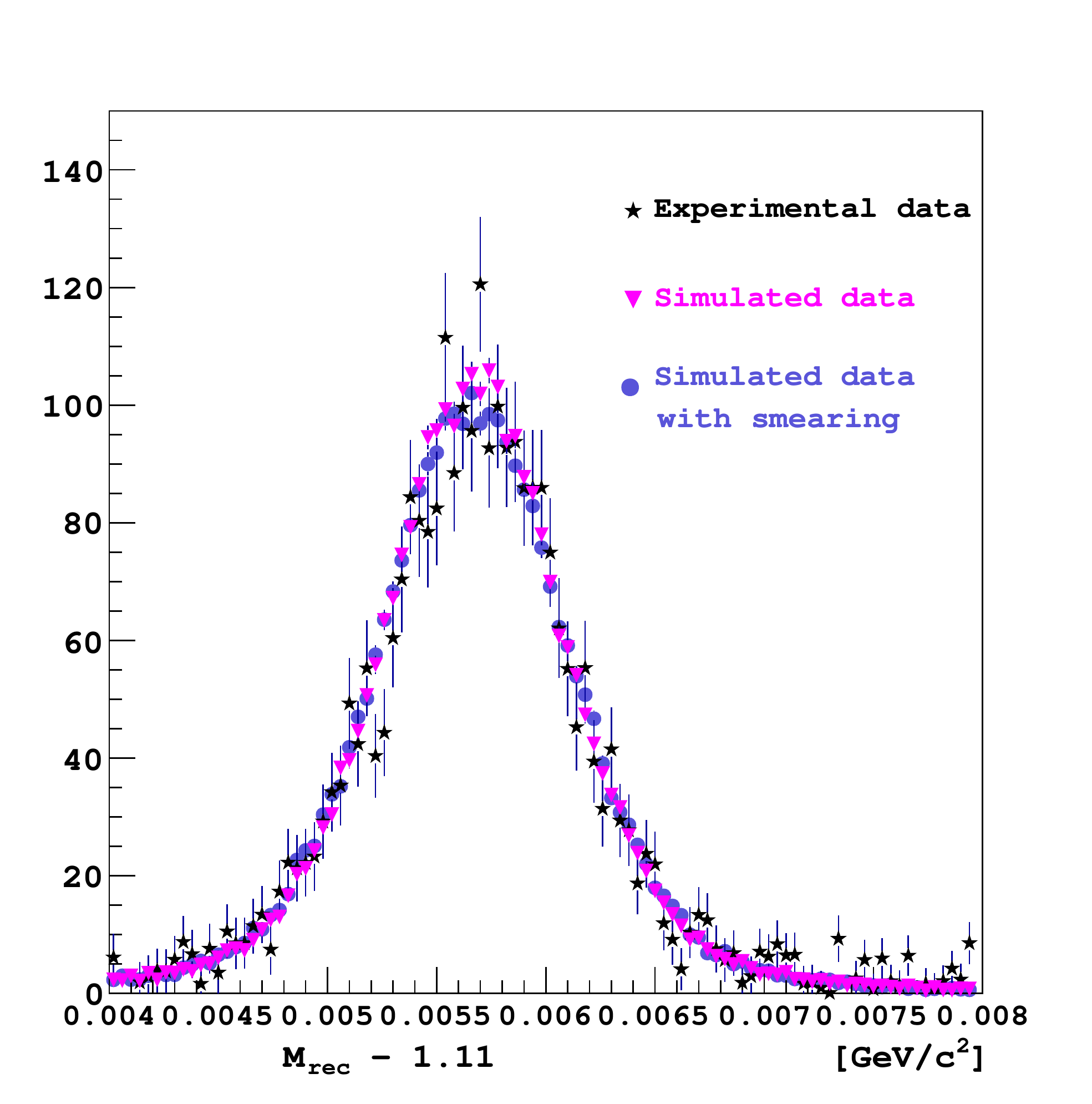}
}
\caption{
Invariant $\pi^-{\rm p}$ mass distribution in the $\Lambda$ region.  
}
\label{Fig3_1}
\end{figure}
The setup has been aligned using properties of $\Lambda$
($\bar{\Lambda}$) decays \cite{note1303}. The decay particles 
are reconstructed and the corresponding invariant 
mass determined. The introduced permanent 
magnet (Fig.~\ref{Fig2_1}) implies a bias in 
the $Y$-components of the charged 
decay particle momenta. The bias amplitude depends on 
the decay vertex position inside the magnet. 
One observes a shift of the mass peak position and 
an increase in the width of the mass distribution. Therefore, 
data with magnetic field only allows to check the equality of 
the $\Lambda$ and $\bar{\Lambda}$ masses. In order to check 
the $\Lambda$ mass value, a special data set without magnet 
has been collected (about 7\% of the total data set). 
The distribution of reconstructed $\Lambda$ masses is 
presented in Fig.~\ref{Fig3_1}. 
The experimental value 
$M_{\Lambda}=1.11568\pm0.00001$~GeV/$c^2$   
is in good agreement  with the PDG value 
$M_{\Lambda}^\mathrm{PDG}=1.115683\pm0.000006$~GeV/$c^2$.    
Investigations of data collected in 2008--2010 \cite{note1303} 
have shown that the width of the simulated $\Lambda$ distribution 
is narrower than the experimental one, due to underestimation of 
the uncertainty in particle momentum reconstruction. 
To improve simulation, a Gaussian smearing of reconstructed momenta 
has been introduced: 
$p^\text{smeared}=p (1 + C_f \cdot N(0,1))$. This smearing of 
simulated momenta with $C_f=(7 \pm 4) \cdot 10^{-4}$ leads to 
a reconstructed $\Lambda$ width consistent with experimental data 
(Fig.~\ref{Fig3_1}). 
The momentum resolution has been evaluated by investigating 
simulated events. For each track the reconstructed 
momentum between 1.5 and 8~GeV/$c$ is compared with 
the generated one. The relative precision lies 
within $2.8\cdot10^{-3}$ to $4.4\cdot10^{-3}$. 
The resolutions in the components of the relative 
pair momentum $Q$, without taking into account 
target multiple scattering, are 
$\sigma(Q_X)=\sigma(Q_Y)=0.44$~MeV/$c$ 
and $\sigma(Q_L)=0.50$~MeV/$c$.

\section{Simulation}
\label{sec:simulation}

A generator, called DIPGEN \cite{note0711}, is used to simulate $\pi^+\pi^-$ pairs 
(atomic pairs, Coulomb and non-Coulomb pairs), which are generated in the Be target and Pt foil. 
For long-lived atoms the following quantities are evaluated:\\
1. Ratio $\epsilon_{nlm}(Be)$ between number $N_A^{L,Be}$ of long-lived atoms, 
which exit the Be target, and total number $N_A$ of produced atoms as a function of 
atom quantum numbers $n, l, m$: $N_A^{L,Be}(n,l,m) = \epsilon_{nlm}(Be) \times N_A$;\\
2. Atom decay probabilities in the space between Be and Pt as 
a function of $n$, $l$  and atom momentum \cite{note1501};\\
3. Ratio $\epsilon_{nlm}(Pt)$ between number $N_A^{L,Pt}$ of long-lived atoms 
at the Pt foil entry and total number of produced atoms: 
$N_A^{L,Pt}(n,l,m) = \epsilon_{nlm}(Pt) \times N_A$;\\ 
4. Breakup probability $W_\mathrm{br}$ in Pt as a function of $n$, $l$ and $m$.\\
In Table~\ref{tab:n}, the relative population of long-lived atom states as a function 
of $n$ is given at the exit of the Be target and at the entry of the Pt foil 
taking into account their decay in-between. 

\begin{table}[htb]
  \caption{Relative populations $\epsilon_{n}(Be)$ and $\epsilon_{n}(Pt)$   
           as a function of n (summed over l and m). 
   }
  \label{tab:n}
  \centering
\begin{tabular}{|c|lllll|}
\hline
$n$            &   2 &   3 &   4 &   5 &  $\geq 2$ \\
\hline
 $\epsilon_{n}(Be) \times 10^2$ & $2.48 \pm O(10^{-3})$ & $1.54 \pm 0.01$ & $0.86 \pm 0.03$ &
 $0.56 \pm 0.06$ & $6.8 \pm 0.6$ \\
 $\epsilon_{n}(Pt) \times 10^2$ & $0.52 \pm O(10^{-4})$ & $1.10 \pm O(10^{-3})$ & $0.78 \pm 0.03$ &
 $0.54 \pm 0.06$ & $4.3 \pm 0.6$ \\
\hline
\end{tabular}
\end{table}

The calculated number of long-lived atoms at the exit of the Be target is 
$N_A^{L,Be}=(6.8 \pm 0.6) \cdot 10^{-2}\times N_A$. After passing 
the gap of 96~mm, $(4.3 \pm 0.6)\%$ of the produced atoms enter the Pt foil.  
Their breakup probability in Pt varies in a wide range depending on
the atom $\left|nlm\right>$ state and momentum at the Pt foil entry: 
e.g. for an input 2p state with 4.5~GeV/$c$, it is about 0.7, 
while for a 5p state the breakup probability exceeds 0.95. 
As a result, the number of atomic pairs generated in Pt  
is $n_A^{L,calc}=(4.0 \pm 0.7) \cdot 10^{-2} \times N_A$, what corresponds 
to an averaged breakup probability of $W_\mathrm{br}=0.92$. 
The magnetic field between Be and Pt has not been considered. 
Its influence is a decrease of the number of atoms 
at the Pt foil entry \cite{NEME85,NEME01} and will be studied in future. 
The atomic (Pt), Coulomb (Be) and non-Coulomb (Be) pair distributions of 
relative momentum projections $q_x$, $q_y$, $q_L$ (initial, i.e. not smeared) 
at the breakup point as well as after multiple scattering have been simulated.
In the next step, all pairs generated by DIPGEN are transferred 
to the GEANT-DIRAC (setup simulator) and ARIANE (reconstruction tool) 
programs. The distributions of the reconstructed values $Q_L$ 
and $Q_T$ of pairs from $A_{2\pi}^L$ breakup in Pt as well as of Coulomb 
and non-Coulomb pairs, generated in Be, are obtained. The majority of 
the atomic pairs from the breakup in Pt has a $Q_T$ and $Q_L$ of less 
than 1.5~MeV/$c$ (Fig.~\ref{Fig5_1} and \ref{Fig5_2}). Moving in 
the fringing magnetic field, each atomic pair receives 
an additional $\Delta Q_Y=2.3$~MeV/$c$. To get the initial value 
of $Q_Y$ for atomic pairs from Pt, the reconstructed value must be 
reduced by this amount, leading to the initial value of the 
transverse momentum $Q_T=\sqrt{Q_X^2+(Q_Y-2.3\:\mathrm{MeV/}c)^2}$.

\section{Data analysis and results}
\label{sec:analysis}

The experimental distributions of $\pi^+\pi^-$ pairs as 
a function of relative momentum $Q$ components have been  
fitted with simulated distributions of atomic pairs ($n_A^L$)  
from Pt, Coulomb ($N_C$) and non-Coulomb pairs ($N_{nC}$) from Be. 
The three corresponding numbers are free parameters in the fit. 

\begin{figure}
\includegraphics[width=\columnwidth]{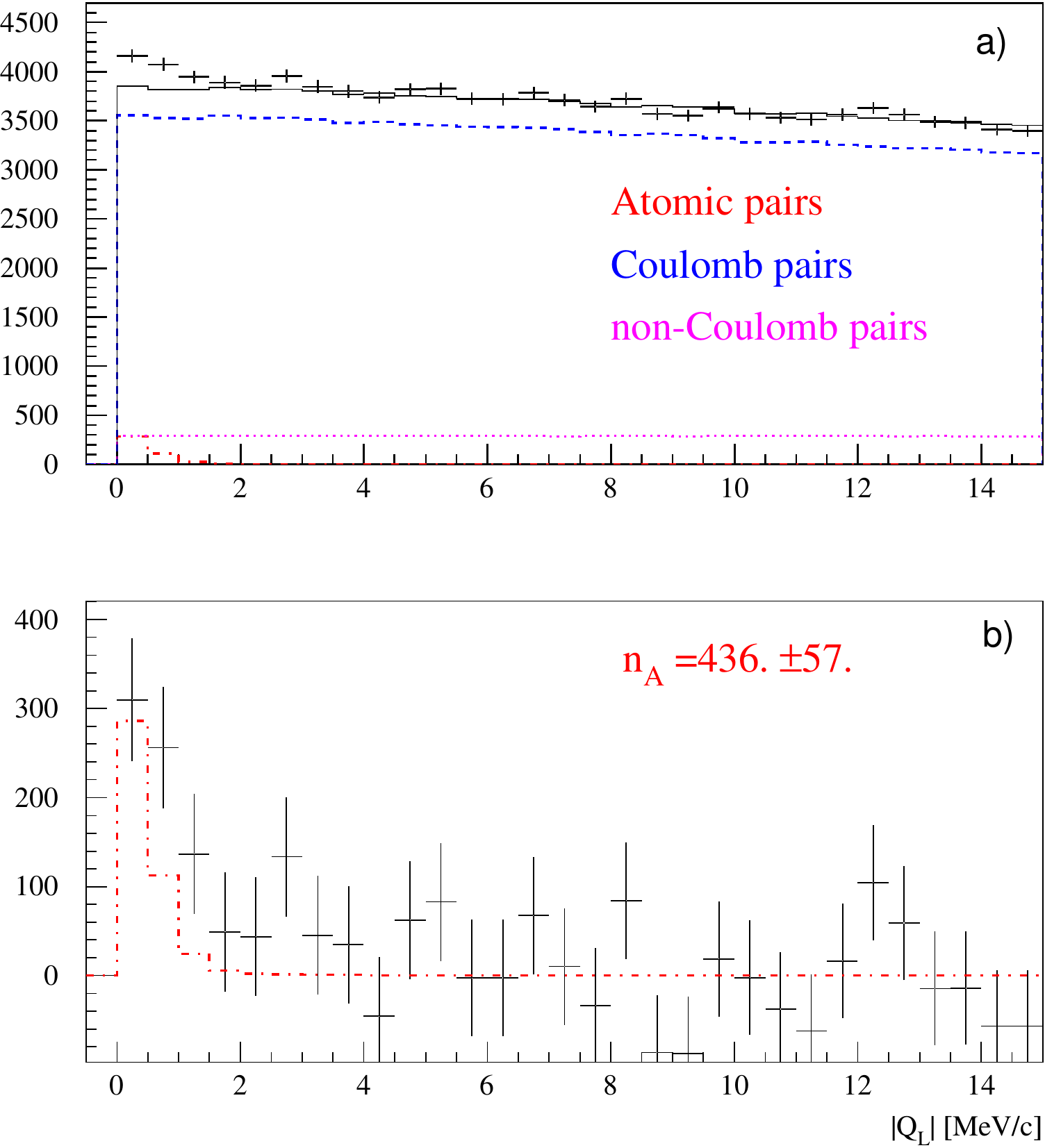}
\caption{$|Q_L|$ distribution of $\pi^+\pi^-$ pairs for 
$Q_T<2.0$~MeV/$c$. The plot a) shows the experimental distribution 
(points with statistical error) and the simulated background (solid line).
The plot b) shows the experimental distribution after 
background subtraction (points with statistical error) and 
the simulated distribution of atomic pairs (dotted-dashed line). 
The fit procedure has been applied to 
the 2-dimensional ($|Q_L|,Q_T$) distribution.}
\label{Fig5_1}
\end{figure}

\begin{figure}
\includegraphics[width=\columnwidth]{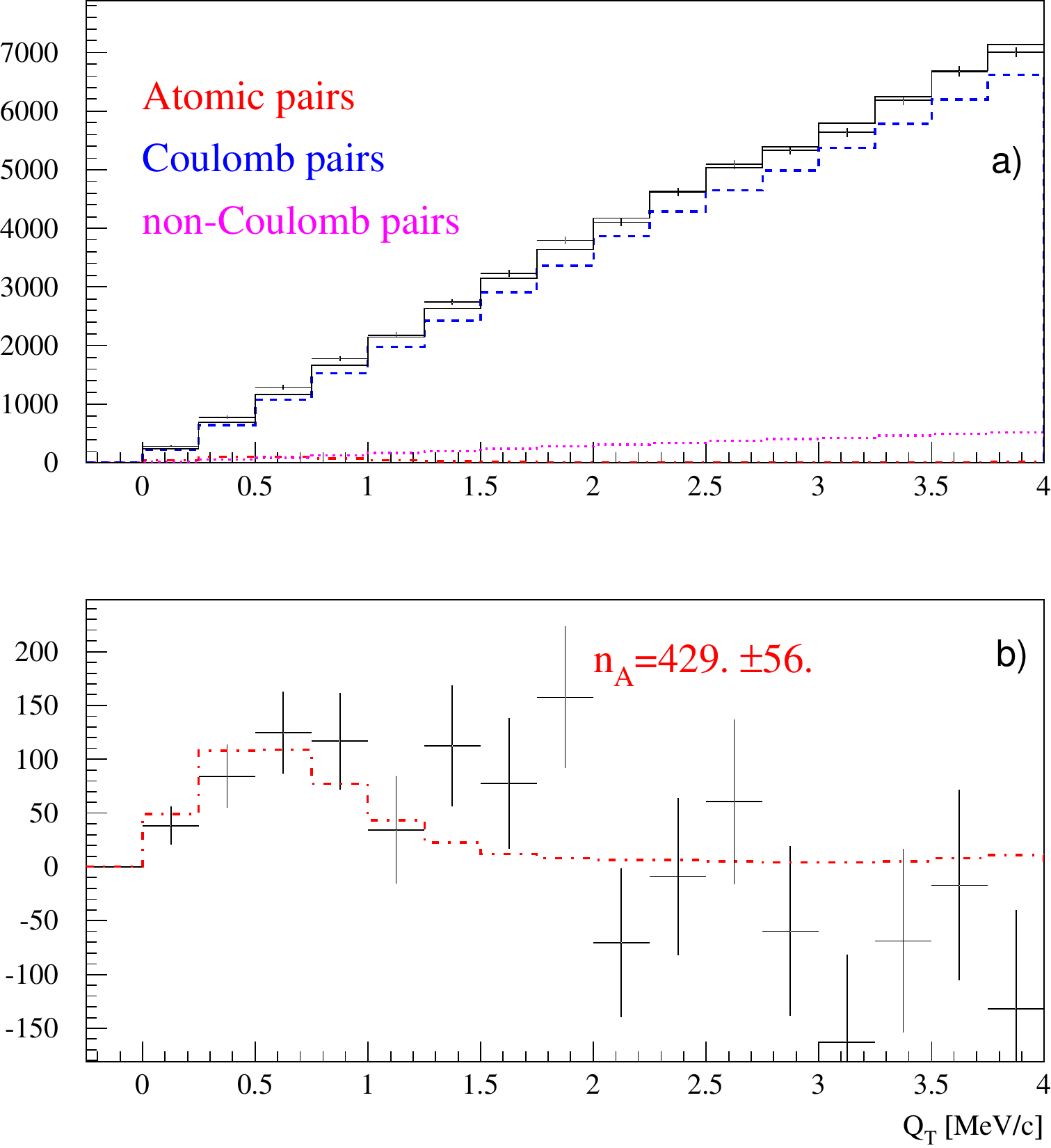}
\caption{$Q_T$ distribution of $\pi^+\pi^-$ pairs for $|Q_L|<2$~MeV/$c$. 
The plot a) shows the experimental distribution (points with 
statistical error) and the simulated background (solid line).
The plot b) shows the experimental distribution after 
background subtraction (points with statistical error) and 
the simulated distribution of atomic pairs (dotted-dashed line). 
The fit procedure has been applied to 
the 2-dimensional ($|Q_L|,Q_T$) distribution.}	
\label{Fig5_2}
\end{figure}

In the 2-dimensional ($|Q_L|,Q_T$) analysis, the experimental data 
have been analysed using simulated 2-dimensional distributions. 
For $|Q_L|<15$~MeV/$c$ and $Q_T<2$~MeV/$c$, 
the $|Q_L|$ projection of the experimental 2-dimensional distribution 
as well as of the three types of simulated $\pi^+\pi^-$ pairs are 
shown in Fig.~\ref{Fig5_1}a. One observes an excess of 
events - above the sum of Coulomb and non-Coulomb pairs - in 
the low $Q_L$ region, where atomic pairs are expected. 
After subtracting background, there is a statistically 
significant signal of $n_A^L=436 \pm 57$  
(Fig.~\ref{Fig5_1}b). The signal shape is described 
by the simulated distribution of atomic pairs 
resulting from the long-lived atom breakup in the Pt foil. 
The atomic pair selection efficiency 
for different cuts on $Q_T$ is known from simulation. 
Using this efficiency the total number of atomic pairs 
generated in Pt is $n_A^{L,tot}=488 \pm 64$. 
Fig.~\ref{Fig5_2}a presents the $Q_T$ projection of 
the same 2-dimensional distributions for 
$|Q_L|<2$~MeV/$c$ and $Q_T<4$~MeV/$c$ with the same free fit parameters. 
After background subtraction in the low $Q$ region, 
one observes again a statistically 
significant signal with a shape described by 
the simulated $Q_T$ distribution of atomic pairs from 
long-lived atom breakup in Pt (Fig.~\ref{Fig5_2}b). 
The number of atomic pairs in the region $Q_T<4$~MeV/$c$ is 
$n_A^L=429 \pm 56$. 

In the 1-dimensional analysis, the $|Q_L|$ distribution  
is fitted using different cuts $Q_T<0.5$, 1.0, 1.5, 2.0~MeV/$c$ 
to study the stability of the atomic pair number for different  
background levels. The detected numbers $n_A^L$ of atomic pairs 
and the corresponding total numbers are shown in Table~\ref{tab:fit}. 
The background for $Q_T<2$~MeV/$c$ is 17 times higher than 
for $Q_T<0.5$~MeV/$c$. Nevertheless, the values in 
the 1- and 2-dimensional analyses coincide within statistics. 
This confirms the signal stability for different $Q_T$ cuts, 
i.e. for different background levels.  

\begin{table}[htb]
  \caption{
Analysis of data collected in 2012 for different fitting procedures. 
The detected numbers $n_A^L$ of atomic pairs and the corresponding 
total numbers $n_A^{L,tot}$ (via selection efficiency) are presented 
together with the background contribution (Coulomb, non-Coulomb and 
accidental pairs) and the fit quality $\chi^{2}/n$ (n = degrees of 
freedom). Errors are only statistical.
}
\label{tab:fit} 
\centering
\begin{tabular}{|c|c|c|r|c|}
\hline
 & & & & \\	
 $Q_T$ cut & $n_A^L$ & $n^{L,tot}_A$ &Back- & $\chi^2$/n \\
 (MeV/$c$) &         &                &ground& \\
\hline
\multicolumn{5}{|c|}{2-dimensional fit over $Q_L,Q_T$} \\
\hline
 & & & & \\
 2.0 & $436 \pm  57$ & $488 \pm 64$ & 16790 & 138/140 \\
 & & & & \\
\hline
\multicolumn{5}{|c|}{1-dimensional fit over $Q_L$} \\
\hline
 & & & & \\
 0.5 & $152 \pm  29$ & $467 \pm 88$  &   971 & 29/27 \\
 1.0 & $349 \pm  53$ & $489 \pm 75$  &  3692 & 19/27 \\
 1.5 & $386 \pm  78$ & $454 \pm 91$  &  9302 & 22/27 \\
 2.0 & $442 \pm 105$ & $495 \pm 117$ & 16774 & 22/27 \\
 & & & & \\
\hline
\end{tabular} \end{table} 

The measurement of the atomic pair number $n_A^L$ is affected by 
the description accuracy of the simulated distributions for 
Coulomb, non-Coulomb and atomic pairs. If the shapes of 
the simulated distributions differ from the experimental ones, 
then the fit parameter values might be biased. In the  
long-lived atom search, the shapes of Coulomb and non-Coulomb pairs 
from Be are similar in phase-space (Fig.~\ref{Fig5_1} and \ref{Fig5_2}). 
Therefore, a systematic error might only arise from 
an incorrect description of the atomic pair distribution, due to  
the uncertainty in the $Q_L$ and $Q_T$ resolution \cite{note1502}. 

There are two main sources of systematic errors in 
the number $n_A^L$ of atomic pairs: 
1) The $\Lambda$ width correction accuracy 
(Section~\ref{sec:track}) leads to 
a systematic error of $\sigma^{syst}_{\Lambda}=4.4$. 
2) The accuracy of the measured Pt foil thickness dominates 
the uncertainty in $Q_T$ resolution, causing  
a systematic error of $\sigma^\mathrm{syst}_{Pt}=22$ 
in the 2-dimensional analysis. In the 1-dimensional $|Q_L|$ analysis, 
this error is nearly 0. 

Another problem might arise from a hypothetical admixture of 
Coulomb pairs generated from beam halo protons interacting with 
the Pt foil. A peak induced by Coulomb final state interaction 
would be at the same place as atomic pairs from long-lived atoms. 
Beam halo level and interaction rate with Pt have been 
investigated \cite{note1302} with the result that the particle flux 
on Pt is practically negligible under working condition. 
To minimize any wrong interpretation, data have also been analysed 
under the assumption that they originate from Coulomb pairs 
generated in Pt and not from long-lived atoms: 
this hypothesis is statistically unlikely \cite{note1502}.

In summary, the 2-dimensional analysis results in 
the following number of atomic pairs from 
$A_{2\pi}^L$  breakup in the Pt foil: $n_A^L=436 \pm 61$. 
This corresponds to 7.1 standard deviations, taking into account 
statistical as well as systematic errors. In order 
to get an estimate for the expected number of atomic pairs,   
pion pairs generated in Be have also been analysed. 
To evaluate the initial value of $Q_Y$ ($Q_T$), 
the momentum shift $\Delta Q_Y$, due to the magnetic field, 
of 12.9~MeV/$c$ has been subtracted from the reconstructed $Q_Y$. 
The 2-dimensional experimental ($|Q_L|,Q_T$) distribution 
has been fitted by simulated distributions of atomic, Coulomb and 
non-Coulomb pairs from Be. Their corresponding numbers $n_A$, 
$N_C$ and $N_{nC}$ are free fit parameters. The total number of 
produced $\pi^+\pi^-$ atoms, $N_A=17043 \pm 410$, has 
been obtained by using the precise (1\%) ratio between 
$N_A$ and $N_C(Q<Q_{cut})$, the number of detected Coulomb pairs 
with small $Q$ \cite{AFAN99}. Knowning from simulation 
(Section~\ref{sec:simulation}) that $(4.0 \pm 0.7)\%$ of 
the produced atoms ($N_A$) break up in the Pt foil and 
the ($|Q_L|,Q_T$) fit selection efficiency is 0.89, 
one estimates a generation of $607 \pm 110$ atomic pairs, 
and this does not contradict the measured number of 
$436 \pm 61$.

\section{Conclusion}
\label{sec:conclusion}

Long-lived $\pi^+\pi^-$ atoms have been observed for the first time 
in a dedicated experiment performed by means of the adapted DIRAC setup. 
Double-exotic $\pi^+\pi^-$ atoms are produced in $ns$ states by 
24~GeV/$c$ CERN PS protons hitting a 103~$\mu$m thick Be target. 
The $\pi^+\pi^-$ pair analysis yields about 17000 
$\pi^+\pi^-$ atoms, based on the measured number of 
small $Q$ Coulomb pairs. These atoms are moving in the target and 
interacting electromagnetically with Be atoms. About 7\% of them 
leave the target in excited long-lived states. 
At a distance of 96~mm downstream of the Be target, 
a 2.1~$\mu$m thick breakup Pt foil has been installed. 
While passing through the gap between Be and Pt, 
some of the bound states (Table~\ref{tab:n}), 
mainly shorter lived states, are decaying, 
whereas the rest enters the Pt foil and 
about 90\% break up, generating $\pi^+\pi^-$ atomic pairs: 
$$
n_A^L = 436 \pm 57|_\mathrm{stat} \pm 23|_\mathrm{syst} = 
436 \pm 61|_\mathrm{tot} \, .
$$ 
This result corresponds to a 7.1$\sigma$ effect  
and does not contradict the estimated value of  
$607\pm110$. 
The observation of long-lived $\pi^+\pi^-$ bound states gives 
the possibility to study the \textit{Lamb shift} and herewith 
a new $\pi\pi$ scattering length combination.

\section*{Acknowledgements}

We are grateful to A.~Vorozhtsov, D.~Tommasini and their colleagues  
from TE-MSC/CERN for the Sm-Co magnet design and construction, 
R.~Steerenberg and the CERN-PS crew for 
the delivery of a high quality proton beam and 
the permanent effort to improve the beam characteristics. 
The project DIRAC has been supported by 
CERN, the JINR administration, the Ministry of Education and 
Youth of the Czech Republic by project LG130131, 
the Istituto Nazionale di Fisica Nucleare and the University of Messina (Italy),  
the Grant-in-Aid for Scientific Research from 
the Japan Society for the Promotion of Science, 
the Ministry of Education and Research (Romania), 
the Ministry of Education and Science of the Russian Federation and 
Russian Foundation for Basic Research, 
the Direcci\'{o}n Xeral de Investigaci\'{o}n, Desenvolvemento e Innovaci\'{o}n, 
Xunta de Galicia (Spain) and the Swiss National Science Foundation.

\end{document}